\documentclass[sigconf]{acmart}

\copyrightyear{2019}
\acmYear{2019}
\setcopyright{licensedothergov}
\acmConference[CASA '19]{Computer Animation and Social Agents}{July 1--3, 2019}{PARIS, France}
\acmBooktitle{Computer Animation and Social Agents (CASA '19), July 1--3, 2019, PARIS, France}
\acmPrice{15.00}
\acmDOI{10.1145/3328756.3328769}
\acmISBN{978-1-4503-7159-9/19/07}

\usepackage{comment}
\usepackage{subcaption}
\usepackage{enumitem}

\newcommand{\SubsectionShort}[1]{\medskip \noindent \textbf{#1.}}

\begin{document}

\title[Data-Driven Crowd Simulation with Generative Adversarial Networks]{Data-Driven Crowd Simulation\\with Generative Adversarial Networks}

\author{Javad Amirian}
\email{javad.amirian@inria.fr}
\affiliation{%
  \institution{Univ Rennes, Inria, CNRS, IRISA}
  \city{Rennes}
  \country{France}
}
\orcid{0003-4484-7134}

\author{Wouter van Toll}
\email{wouter.van-toll@inria.fr}
\affiliation{%
  \institution{Univ Rennes, Inria, CNRS, IRISA}
  \city{Rennes}
  \country{France}
}
\orcid{0003-1225-1272}

\author{Jean-Bernard Hayet}
\email{jbhayet@cimat.mx}
\affiliation{%
  \institution{Centro de Investigaci\'on en Matem\'aticas}
  \city{Guanajuato}
  \country{Mexico}
}
\orcid{0002-6662-2553}

\author{Julien Pettr\'e}
\email{julien.pettre@inria.fr}
\affiliation{%
  \institution{Univ Rennes, Inria, CNRS, IRISA}
  \city{Rennes}
  \country{France}
}
\orcid{0003-1812-1436}

\begin{abstract}
This paper presents a novel data-driven crowd simulation method that can mimic the observed traffic of pedestrians in a given environment. 
Given a set of observed trajectories, 
we use a recent form of neural networks, Generative Adversarial Networks (GANs), to learn the properties of this set and generate new trajectories with similar properties. 
We define a way for simulated pedestrians (agents) to follow such a trajectory while handling local collision avoidance. 
As such, the system can generate a crowd that behaves similarly to observations, while still enabling real-time interactions between agents.
Via experiments with real-world data, we show that our simulated trajectories preserve the statistical properties of their input.
Our method simulates crowds in real time that resemble existing crowds, 
while also allowing insertion of extra agents, combination with other simulation methods, and user interaction.
\end{abstract}



\maketitle

\section{Introduction}\label{s:introduction}

The realistic simulation of human crowd motion is a vast research topic 
that includes aspects of artificial intelligence, computer animation, motion planning, psychology, and more. 
Generally, the goal of a crowd simulation algorithm is to populate a virtual scene with a crowd that exhibits visually convincing behavior. 
The simulation should run in real time to be usable for interactive applications such as games, training software, and virtual-reality experiences. 
Many simulations are \emph{agent-based}: they model each pedestrian as a separate intelligent agent with individual properties and goals.

To simulate complex behavior, \emph{data-driven} crowd simulation methods use real-world input data (such as camera footage) to generate matching crowd motion.
Usually, these methods cannot easily generate \emph{new} behavior that is not literally part of the input. 
Also, they are often difficult to use for applications in which agents need to adjust their motion in real-time, 
e.g.\ because the user is part of the crowd.
\medskip

\noindent
In this paper, we present a new data-driven crowd simulation method that largely avoids these limitations. 
Our system enables the real-time simulation of agents that behave similarly to observations, while allowing them to deviate from their trajectories when needed. 
More specifically, our method:
\begin{enumerate}[leftmargin=*]
\item learns the overall properties of input trajectories, and can generate new trajectories with similar properties;
\item embeds these trajectories in a crowd simulation, in which agents follow a trajectory while allowing for local interactions.
\end{enumerate}

\noindent
For item 1, we use Generative Adversarial Networks (GANs) \cite{Goodfellow2014-GAN}, a novel technique in machine learning for generating new content based on existing data. 
For item 2, we extend the concept of `route following' \cite{Jaklin2013-MIRAN} to trajectories with a \emph{temporal} aspect, 
prescribing a speed that may change over time. 

Using a real-world dataset as an example, we will show that our method generates new trajectories with matching styles. 
Our system can (for example) reproduce an existing scenario with additional agents, and it can easily be combined with other crowd simulation methods. 

\section{Related Work}\label{s:relatedwork}

Agent-based crowd simulation algorithms model pedestrians as individual intelligent agents.
In this paradigm, many researchers focus on the \emph{local} interactions between pedestrians 
(e.g.\ collision avoidance) using \emph{microscopic} algorithms \cite{Helbing1995-SocialForces,vandenBerg2011-ORCA}. 
In environments with obstacles, these need to be combined with \emph{global} path planning into an overall framework \cite{vanToll2015-Framework}. 
A growing research topic lies in measuring the `realism' of a simulation, 
by measuring the similarity between two fragments of (real or simulated) crowd motion \cite{Wang2016-PathPatterns}.
\medskip

\noindent
Complex real-life behavior can hardly be described with simple local rules. 
This motivates \emph{data-driven} simulation methods, 
which base the crowd motion directly on real-world trajectories, typically obtained from video footage. 
One category of such methods stores the input trajectories in a database, 
and then pastes the best-matching fragments into the simulation at run-time \cite{Lerner2007-CrowdsByExample,Lee2007-GroupBehavior}. 
Another technique is to create pre-computed \emph{patches} with periodically repeating crowd motion, 
which can be copied and pasted throughout an environment \cite{Yersin2009-CrowdPatches}. 
Such simulations are computationally cheap, but difficult to adapt to interactive situations.

Researchers have also used input trajectories to train the parameters of (microscopic) simulation models \cite{Wolinski2014-ParameterEstimation}, so as to adapt the agents' local behavior parameters to match the input data. 
However, this cannot capture any complex (social) rules that are not part of the used simulation model.

To replicate how agents move through an environment at a higher level, 
some algorithms subdivide the environment into cells and learn how pedestrians move between them \cite{Pellegrini2012-DestinationFlow,Zhong2016-BehaviorLearning}.
Our goal is similar (reproducing pedestrian motion at the full trajectory level), but our approach is different: we learn the spatial and temporal properties of complete trajectories, generate new trajectories with similar properties, and let agents flexibly follow these trajectories.
\medskip

\noindent
Our work uses Generative Adversarial Networks (GANs) \cite{Goodfellow2014-GAN}, a recent AI development for generating new data. 
GANs have been successful at generating creative content such as faces~\cite{Di2018-FaceSynthesis}.
Recently, researchers have started to adopt GANs for short-term prediction of pedestrian motion
\cite{Amirian2019-SocialWays}. 
To our knowledge, our work is the first to apply GANs in crowd simulation at the full trajectory level.

\section{Generating Trajectories}\label{s:generator}



\newcommand {\Pair}[2] {\ensuremath{\langle #1 , #2 \rangle}}
\newcommand {\SampledSubTrajectory}[2] {\ensuremath{\mathbf{p}^\pi_{ #1 : #2 }}}
\newcommand {\SampledTrajectoryPoint}[1] {\ensuremath{\mathbf{p}^\pi_{#1}}}
\newcommand {\SampledTrajectoryPointTraining}[1] {\ensuremath{\mathbf{q}^\pi_{#1}}}
\newcommand {\GeneratedTrajectoryPoint}[2] {\ensuremath{g(\mathbf{z} | \SampledSubTrajectory{0}{#2}; \theta_{G})}}
\newcommand {\DiscriminationEntry} {\ensuremath{v_e(\SampledSubTrajectory{0}{1}; \theta_{D})}}
\newcommand {\DiscriminationContinuation}[1] {\ensuremath{v_c(\SampledSubTrajectory{0}{#1}; \theta_{D})}}
\begin{figure*}[htb]
	\centering
	\includegraphics[trim={0 200 0 200},clip, width=0.9\textwidth]{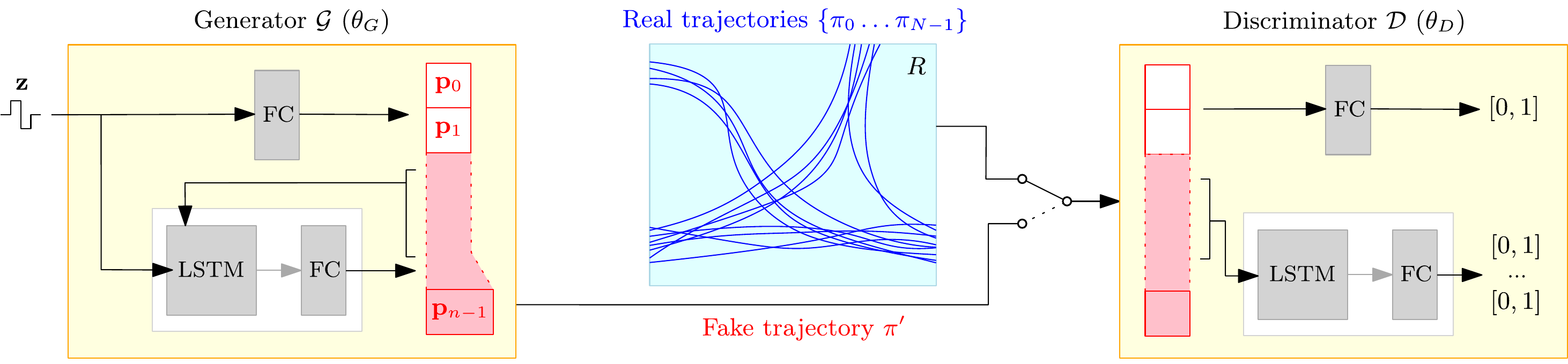}
	\captionsetup{format=plain, font=footnotesize, labelfont=bf}
	\caption{Our GAN architecture for learning and generating pedestrian trajectories.}
	\label{fig:GAN_Arch}
\end{figure*}

In this section, we describe our GAN-based method for generating trajectories that are similar to the examples in our training data.

As in most crowd-simulation research, we assume a planar environment and we model agents as disks.
We define a \emph{trajectory} as a mapping $\pi$: $[0,T] \rightarrow \mathbb{R}^2$ that describes how an agent moves through an environment during a time period of $T$ seconds. 
Note that a trajectory encodes \emph{speed} information: our system should capture when agents speed up, slow down, or stand still.

In practice, we will represent a trajectory $\pi$ by a sequence of $n_\pi$ points 
$[\SampledTrajectoryPoint{0}, \ldots, \SampledTrajectoryPoint{n_{\pi}-1}]$ 
separated by a fixed time interval $\Delta t$; that is, each \SampledTrajectoryPoint{i}\ has a corresponding timestamp $i \cdot \Delta t$. 
In our experiments, we use $\Delta t = 0.4$s because our input data uses this value as well.
We will use the notation \SampledSubTrajectory{j}{k}\ to denote a sub-trajectory from \SampledTrajectoryPoint{j}\ to \SampledTrajectoryPoint{k}.

Given a dataset of trajectories $\Pi = \{\pi_0, \ldots, \pi_{N-1}\}$, 
our generator should learn to produce new trajectories with properties similar to those in $\Pi$. 
We assume that all trajectories start and end on the boundary of a \emph{region of interest} $R$, 
which can have any shape and can be different for each environment. 


\SubsectionShort{Overview of GANs}
A Generative Adversarial Network \cite{Goodfellow2014-GAN} consists of two components: 
a \emph{generator} $\mathcal{G}$ that creates new samples 
and a \emph{discriminator} $\mathcal{D}$ that judges whether a sample is real or generated. 
The training phase of a GAN is a two-player game in which $\mathcal{G}$ learns to `fool' $\mathcal{D}$, 
until (ideally) the generated samples are so convincing that $\mathcal{D}$ does not outperform blind guessing. 

Internally, both $\mathcal{G}$ and $\mathcal{D}$ are artificial neural networks; 
let $\theta_G$ and $\theta_D$ be their respective weights. 
$\mathcal{G}$ acts as a function that converts an $m$-dimensional noise vector $\mathbf{z}$ 
to a fake sample $\mathbf{x} = \mathcal{G}(\mathbf{z}; \theta_{G})$.
$\mathcal{D}$ acts as a function that converts a (real or fake) sample $\mathbf{x}$ 
to a value $\mathcal{D}(\mathbf{x}; \theta_{D}) \in [0,1]$ indicating the probability of $\mathbf{x}$ being real.
Training a GAN represents the following optimization problem:
\begin{equation}
\begin{split}
& \min_{\theta_G} \max_{\theta_D} V(\theta_D, \theta_G), \\
V(\theta_D, \theta_G) = & 
~ \mathbb{E}_{\mathbf{x} \sim p_x}[\log \mathcal{D}(\mathbf{x};\theta_D)] +\\ 
& ~ \mathbb{E}_{\mathbf{z} \sim p_z}[\log(1-\mathcal{D}(\mathcal{G}(\mathbf{z};\theta_G);\theta_D))]
\end{split}
\end{equation}
where $V(\theta_D, \theta_G)$ is known as the \emph{loss function}. 
Its first term denotes the expected output of $\mathcal{D}$ for a random real sample.
This is higher when $\mathcal{D}$ correctly classifies more input samples as real.
Conversely, the second term is higher when $\mathcal{D}$ classifies more generated samples as fake.
Here, $p_x$ and $p_z$ are the probability distributions of (respectively) the real data and the noise vectors sent to $\mathcal{G}$.

\SubsectionShort{Overview of Our System}
Figure~\ref{fig:GAN_Arch} displays an overview of our GAN system. The generator and discriminator both have two tasks: 
generating or evaluating the \emph{entry points} of a trajectory $\pi$ 
(i.e.\ the first two points \SampledTrajectoryPoint{0}\ and \SampledTrajectoryPoint{1}),
and generating or evaluating the \emph{continuation} of a trajectory 
(i.e.\ the next point \SampledTrajectoryPoint{k+1} after a sub-trajectory \SampledSubTrajectory{0}{k}).
For the continuation tasks, we use concepts from so-called `conditional GANs' because the generator and discriminator take extra data as input. 
We will now describe the system in more detail. Parameter settings will be mentioned in Section~\ref{s:experiments}.

\SubsectionShort{Generator}
To generate \emph{entry points}, the generator $\mathcal{G}$ feeds a random vector $\mathbf{z}$ to a fully connected (FC) block of neurons. 
Its output is a 4D vector that contains the coordinates of \SampledTrajectoryPoint{0}\ and \SampledTrajectoryPoint{1}. 

To generate the \emph{continuation} of a trajectory \SampledSubTrajectory{0}{k}, 
the generator $\mathcal{G}$ feeds \SampledSubTrajectory{0}{k}\ and a noise vector $\mathbf{z}$ 
to a Long Short Term Memory (LSTM) layer that should encode the relevant trajectory dynamics.
LSTMs are common \emph{recurrent} neural networks used for handling sequential data.
The output of this LSTM block is sent to an FC block, 
which finally produces a 2D vector with the coordinates of \SampledTrajectoryPoint{k+1}. 
Let \GeneratedTrajectoryPoint{k+1}{k} denote this result.
Ideally, this point will be taken from the (unknown) distribution of likely follow-ups for \SampledSubTrajectory{0}{k}.

The continuation step is repeated iteratively. 
If the newly generated point \SampledTrajectoryPoint{k+1}\ lies outside of the region of interest $R$, then the trajectory is considered to be finished. 
Otherwise, the process is repeated with inputs \SampledSubTrajectory{0}{k+1}\ and a new noise vector. 

\SubsectionShort{Discriminator}
The discriminator $\mathcal{D}$ takes an entire (real or fake) trajectory $\pi$ as input. 
It splits the discrimination into two tasks with a similar structure as in $\mathcal{G}$.
For the \emph{entry point} part, an FC block evaluates \SampledSubTrajectory{0}{1} to a scalar in $[0,1]$, which we denote by \DiscriminationEntry.
For the \emph{continuation} part, 
an LSTM+FC block separately evaluates each point \SampledTrajectoryPoint{k}\ (for $2 \leq k < n_\pi$) 
given the sub-trajectory \SampledSubTrajectory{0}{k-1}. 
We denote the result for the $k$th point by \DiscriminationContinuation{k}.

So, for a full trajectory $\pi$ of $n_\pi$ points, the discriminator computes $n_\pi-1$ scalars that together 
indicate the likelihood of $\pi$ being real. 
The training phase uses these numbers in its loss function.


\SubsectionShort{Training} 
Each training iteration lets $\mathcal{G}$ generate a set $\Pi'$ of $N$ trajectories for different (sequences of) noise vectors.
We then let $\mathcal{D}$ classify all trajectories (both real and fake). 
The loss function of our GAN is the sum of two components:
\begin{itemize}[leftmargin=*]
\item the success rate for discriminating all entry points:
\begin{equation*}
\sum_{\pi \in \Pi} \log \DiscriminationEntry +
\sum_{\pi \in \Pi'} \log(1-\DiscriminationEntry),
\end{equation*}	
\item the success rate for discriminating all other points:
\begin{equation*}
\sum_{\pi \in \Pi} \sum_{k=2}^{n_{\pi}-1} \log \DiscriminationContinuation{k} + 
\sum_{\pi \in \Pi'} \sum_{k=2}^{n_{\pi}-1} \log (1 - \DiscriminationContinuation{k}).
\end{equation*}
\end{itemize}
To let our GAN train faster, we add a third component. 
For each real trajectory $\pi \in \Pi$, we take all valid sub-trajectories \SampledSubTrajectory{k}{k+4} of length $5$
and let $\mathcal{G}$ generate its own version of \SampledTrajectoryPoint{k+4} given \SampledSubTrajectory{k}{k+3}.
We add to our loss function:
\begin{equation*}
\sum_{\pi \in \Pi} \sum_{k=0}^{n_{\pi}-5} || \SampledTrajectoryPoint{k+4} - \GeneratedTrajectoryPoint{k+4}{k+3} ||
\end{equation*}
i.e.\ we sum up the Euclidean distances between real and generated points.
We observed that this additional component leads to much faster convergence and better back-propagation.

To reduce the chance of `mode collapse' (i.e.\ convergence to a limited range of samples), 
we use an `unrolled' GAN~\cite{Metz2017-UnrolledGAN}. 
This is an extended GAN where each optimization step for $\theta_G$ uses an improved version of the discriminator 
that is $u$ steps further ahead (where $u$ is a parameter).

\section{Crowd Simulation}\label{s:crowdsimulation}

Recall that our goal is to use our trajectories in a real-time interactive crowd simulation, 
where agents should be free to deviate from their trajectories if needed. 
This section describes how we combine our trajectory generator with a crowd simulator.

Our approach fits in the paradigm of multi-level crowd simulation \cite{vanToll2015-Framework}, 
in which global planning (i.e.\ computing trajectories) is detached from the simulation loop. 
This loop consists of discrete timesteps. In each step, new agents might be added, 
and each agent tries to follow its trajectory while avoiding collisions.


\SubsectionShort{Adding Agents} 
To determine when a new agent should be added to the simulation, 
we use an exponential distribution whose parameter $\lambda$ denotes the average time between two insertions. 
This parameter can be obtained from an input dataset (to produce similar crowdedness), 
but one may also choose another value deliberately. 
Each added agent follows its own trajectory produced by our GAN.


\SubsectionShort{Trajectory Following}
In each frame of the simulation loop, each agent should try to proceed along its trajectory $\pi$ while avoiding collisions. 
The main difference with classical `route following' \cite{Jaklin2013-MIRAN} 
is that our trajectories have a \emph{temporal} component: they prescribe at what speed an agent should move, and this speed may change over time.
Therefore, we present a way to let an agent flexibly follow $\pi$ while respecting its spatial \emph{and} temporal data.
Our algorithm computes a \emph{preferred velocity} $v_\text{pref}$ that would send the agent farther along $\pi$.
This $v_\text{pref}$ can then be fed to any existing collision-avoidance algorithm, 
to compute a velocity that is close to $v_\text{pref}$ while avoiding collisions with other agents. 

Two parameters define how an agent follows $\pi$:
the \emph{time window} $w$ and the \emph{maximum speed} $s_\text{max}$. 
An agent always tries to move to a point that lies $w$ seconds ahead along $\pi$, taking $s_\text{max}$ into account. 
During the simulation, let $t$ be the time that has passed since the agent's insertion. 
Ideally, the agent should have reached $\pi(t)$ by now.
Our algorithm consists of the following steps:

\begin{enumerate}[leftmargin=*]
\item Compute the \emph{attraction point} $\mathbf p_\text{att} = \pi(t_\text{att})$, where $t_\text{att} = \min(t+w, T)$  
and $T$ is the end time of $\pi$. 
Thus, $\mathbf p_\text{att}$ is the point that lies $w$ seconds ahead of $\pi(t)$, clamped to the end of $\pi$ if needed. 
\item Compute the \emph{preferred velocity} $\mathbf v_\text{pref}$ as $\frac{\mathbf p_\text{att} - \mathbf p}{t_\text{att} - t}$, 
where $\mathbf p$ is the agent's current position. 
Thus, $\mathbf v_\text{pref}$ is the velocity that will send the agent to $\mathbf p_\text{att}$, with a speed based on the difference between $t$ and $t_\text{att}$. 
\item If $||\mathbf v_\text{pref}|| > s_\text{max}$, scale $\mathbf v_\text{pref}$ so that $||\mathbf v_\text{pref}|| = s_\text{max}$. 
This prevents the agent from receiving a very high speed after it has been blocked for a long time.
\end{enumerate}

\noindent \textbf{Collision Avoidance.} 
The preferred velocity $\mathbf v_\text{pref}$ computed by our algorithm can be used as input for any collision-avoidance routine.
In our implementation, we use the popular ORCA method \cite{vandenBerg2011-ORCA}.
In preliminary tests, other methods such as social forces \cite{Helbing1995-SocialForces} proved to be less suitable for our purpose.

\section{Experiments and Results}\label{s:experiments}

\noindent
\textbf{Set-up.} 
We have implemented our GAN using the \emph{PyTorch} library (\url{https://pytorch.org/}).
The input noise vectors are 3-dimensional and drawn from a uniformly random distribution. 
In both $\mathcal{G}$ and $\mathcal{D}$, 
the entry-point FC blocks consist of 3 layers with 128, 64, and 32 hidden neurons, respectively.
For the continuation part, the LSTM blocks consist of 62 cells, and the FC blocks contain 2 layers of 64 and 32 hidden neurons.
To save time and memory, the LSTM blocks only consider the last 4 samples of a sub-trajectory.

For training the GAN, all FC layers use \emph{Leaky-ReLU} activation functions (with negative slope $0.1$), 
to let the gradient always back-propagate, which avoids gradient-vanishing issues. 
We train the GAN for $50{,}000$ iterations, using an unrolling parameter $u = 10$. 

In the crowd simulation, we model agents as disks with a radius of $0.25$m, 
and we use a simulation loop with a fixed frame length of $0.1$s. 
In each frame, all agents perform our route-following algorithm (with $w = 5$ and $s_\text{max} = 2$m/s), 
followed by the ORCA algorithm~\cite{vandenBerg2011-ORCA} as implemented by the original authors.
We remove an agent when it reaches the end of its trajectory.

We test our method on the \emph{ETH} dataset \cite{Pellegrini2009-SocialTracking} that contains recorded trajectories around the entrance of a university building.
We have defined the region of interest $R$ as an axis-aligned bounding box, 
and we use only the 241 trajectories that both enter and exit $R$. 

\begin{figure}[t!]
	\vspace{-0.4cm}
	\begin{subfigure}{\columnwidth}
		\centering
		\includegraphics[trim= 60 382 60 0, clip, width=\columnwidth]{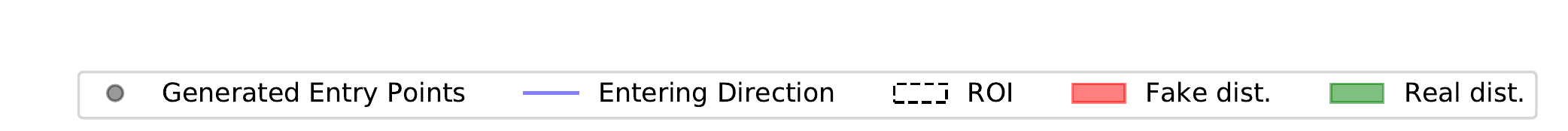}
	\end{subfigure}\\

	\begin{subfigure}{.33\columnwidth}
		\centering
		\includegraphics[trim= 40 40 40 40, clip, width=0.97\columnwidth]{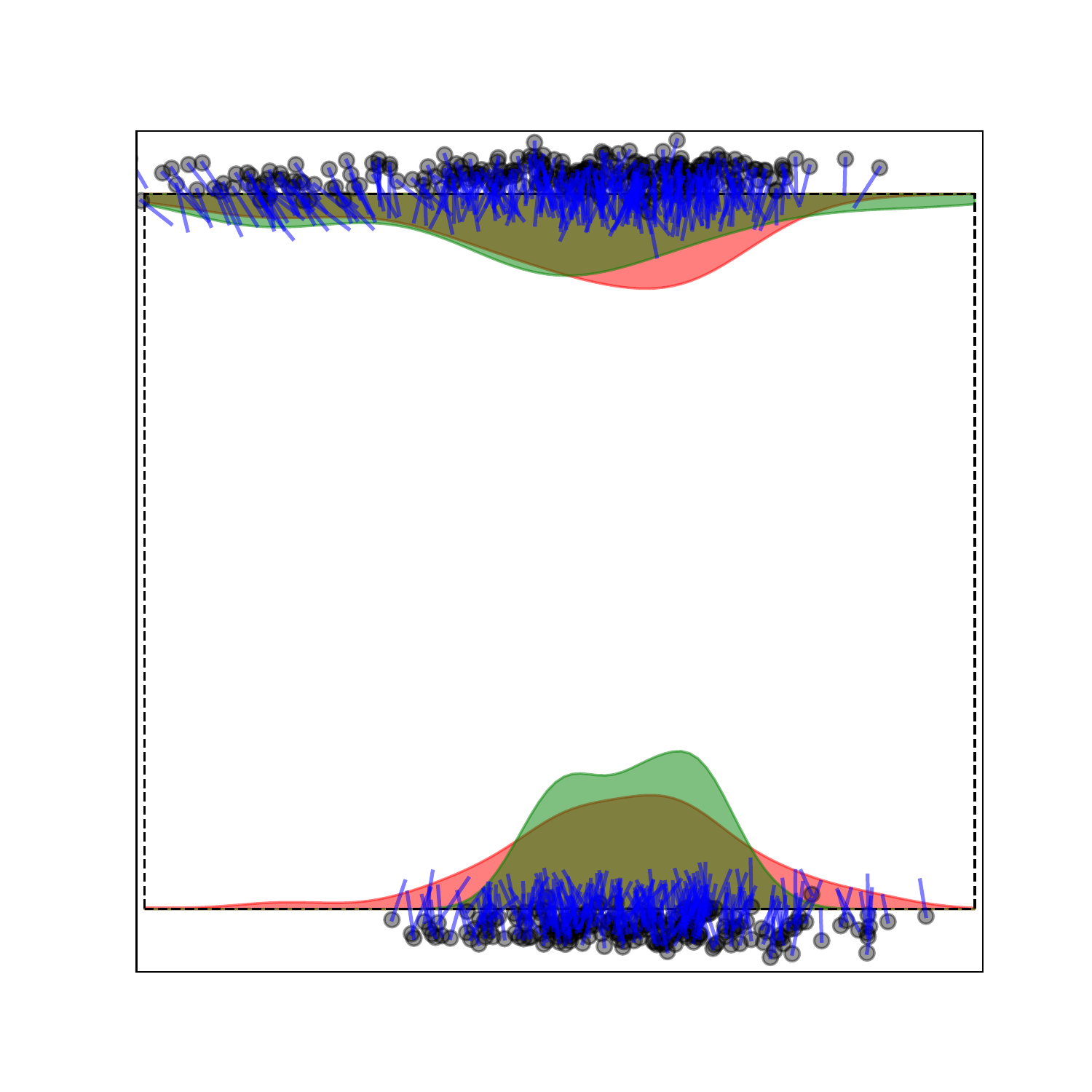}
		\caption{GMM}
	\end{subfigure}%
	\begin{subfigure}{.33\columnwidth}
		\centering
		\includegraphics[trim= 40 40 40 40, clip, width=0.97\columnwidth]{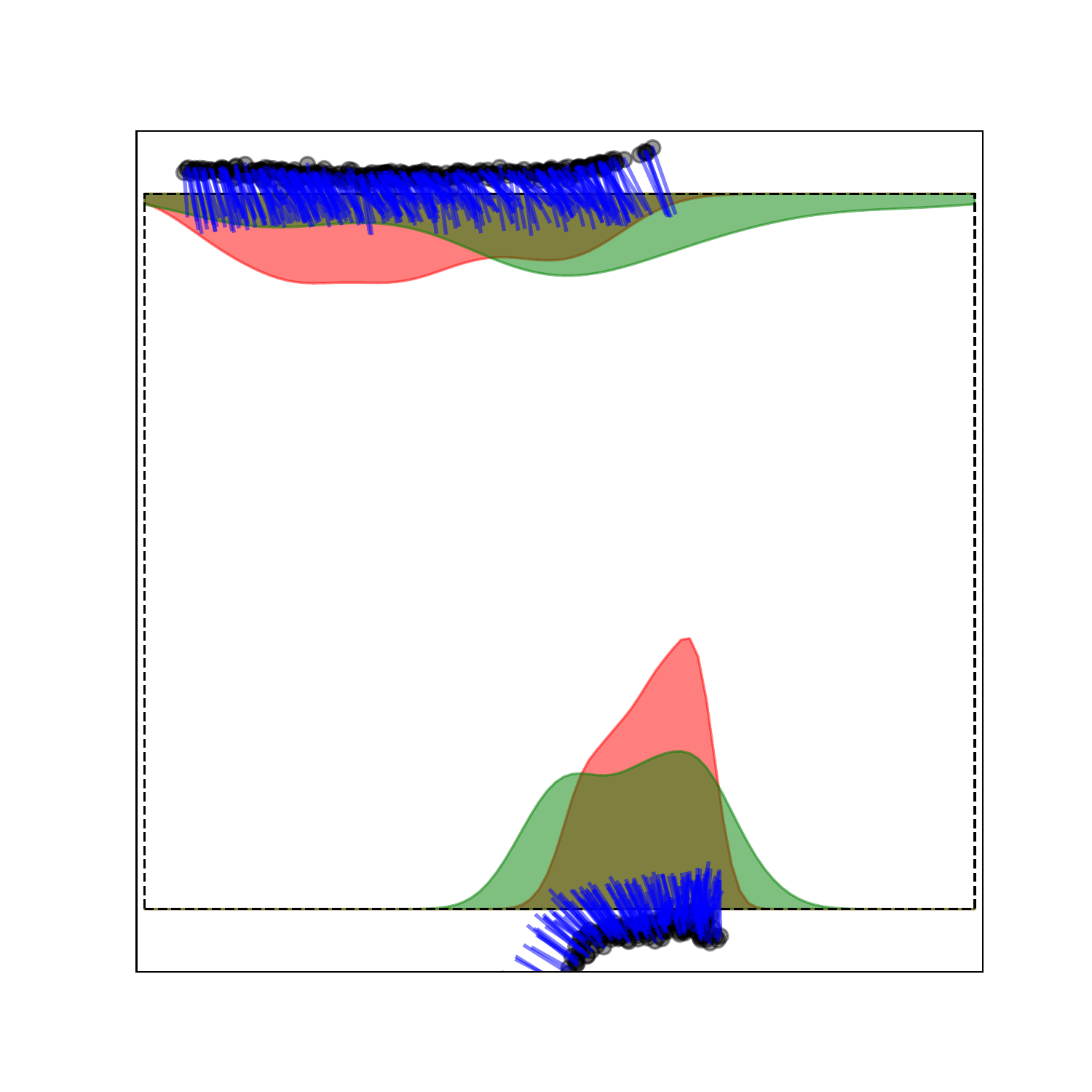}
		\caption{Vanilla GAN}
	\end{subfigure}%
	\begin{subfigure}{.33\columnwidth}
		\centering
		\includegraphics[trim= 40 40 40 40, clip, width=0.97\columnwidth]{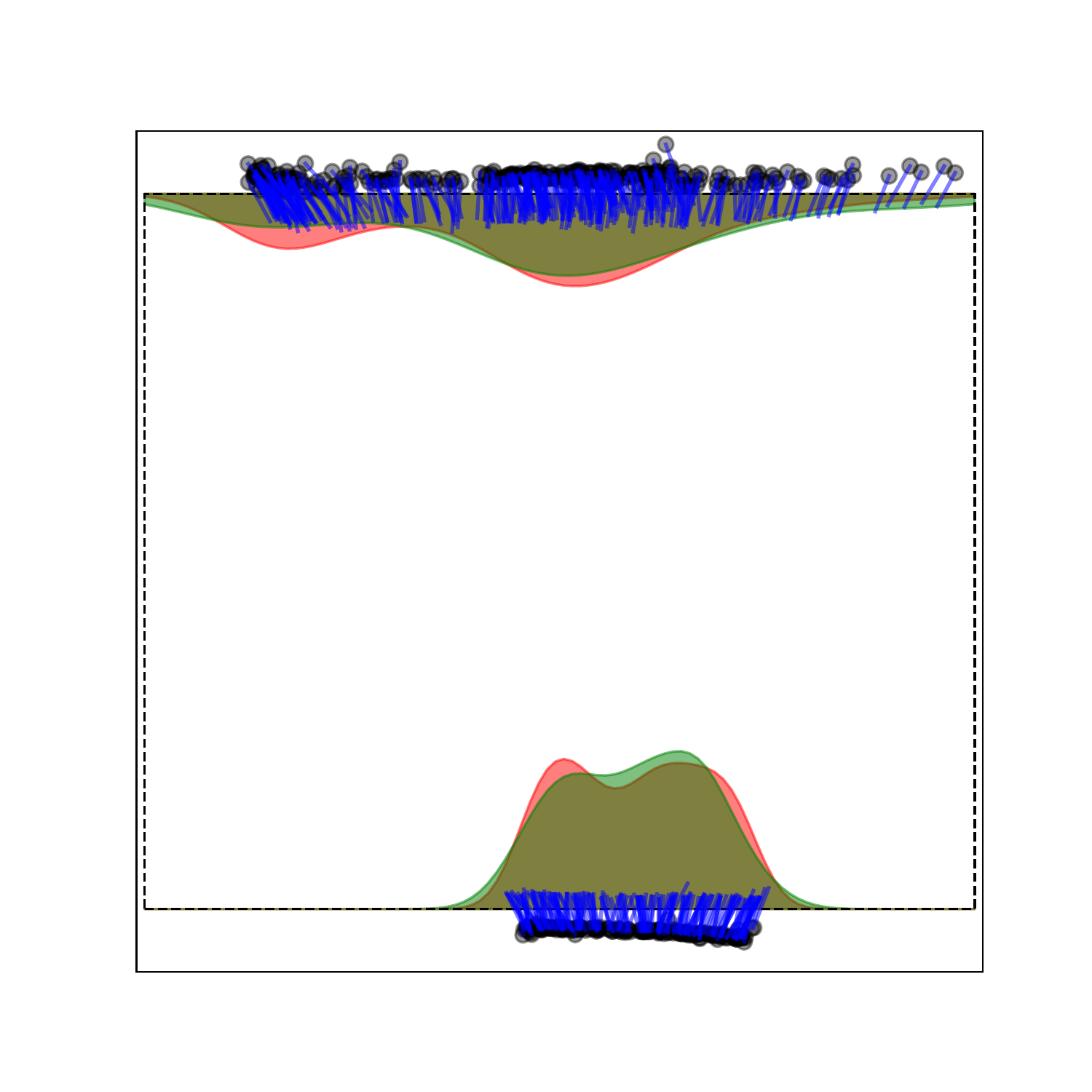}
		\caption{Unrolled GAN}
	\end{subfigure}
	\vspace{-0.2cm}
	\captionsetup{format=plain, font=footnotesize, labelfont=bf}
	\caption{The distribution of entry points created by three different methods.}
	\label{fig:figEntryPoints}
\end{figure}

\SubsectionShort{Result 1: Entry Points} 
To show the performance of our GAN in learning the distribution of entry points, we computed 500 (fake) entry points in the ETH scene, and we calculated the distribution of the samples over the boundary of $R$. 
We also compared these results against two other generative methods: 
a Gaussian Mixture Model (GMM) with 3 components, and a `vanilla' GAN variant that does not use the unrolling mechanism. 
As shown in Fig.~\ref{fig:figEntryPoints}, the entry points of the unrolled GAN (right) are closer to the real data than those of the other two methods.

\begin{figure}[t!]
	\begin{subfigure}{.33\columnwidth}
		\centering
		\includegraphics[trim= 280 100 280 200, clip, width=0.97\columnwidth] {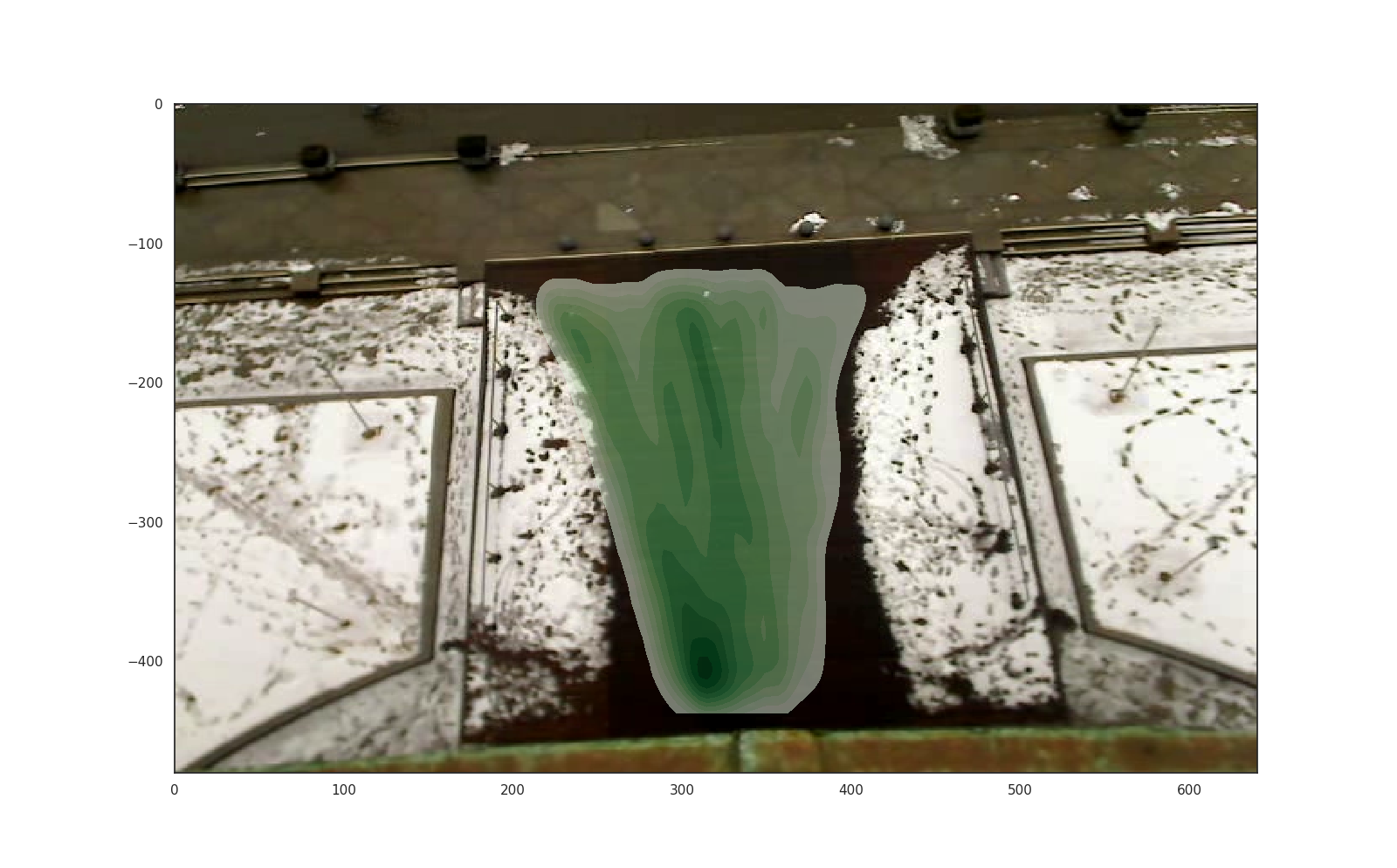}
	\end{subfigure}%
	\begin{subfigure}{.33\columnwidth}
		\centering
		\includegraphics[trim= 280 100 280 200, clip, width=0.97\columnwidth] {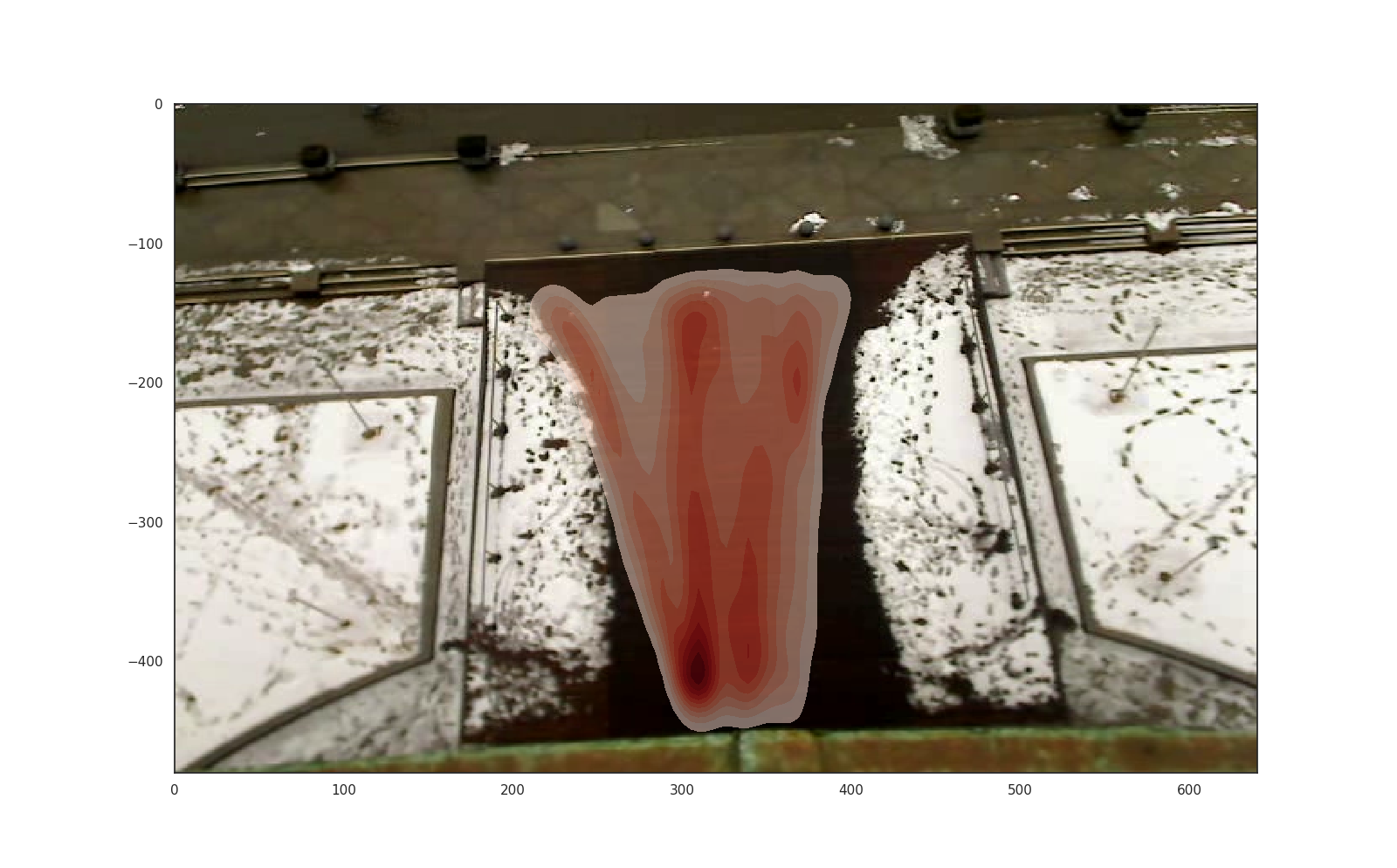}
	\end{subfigure}%
	\begin{subfigure}{.33\columnwidth}
		\centering
		\includegraphics[trim= 280 100 280 200, clip, width=0.97\columnwidth] {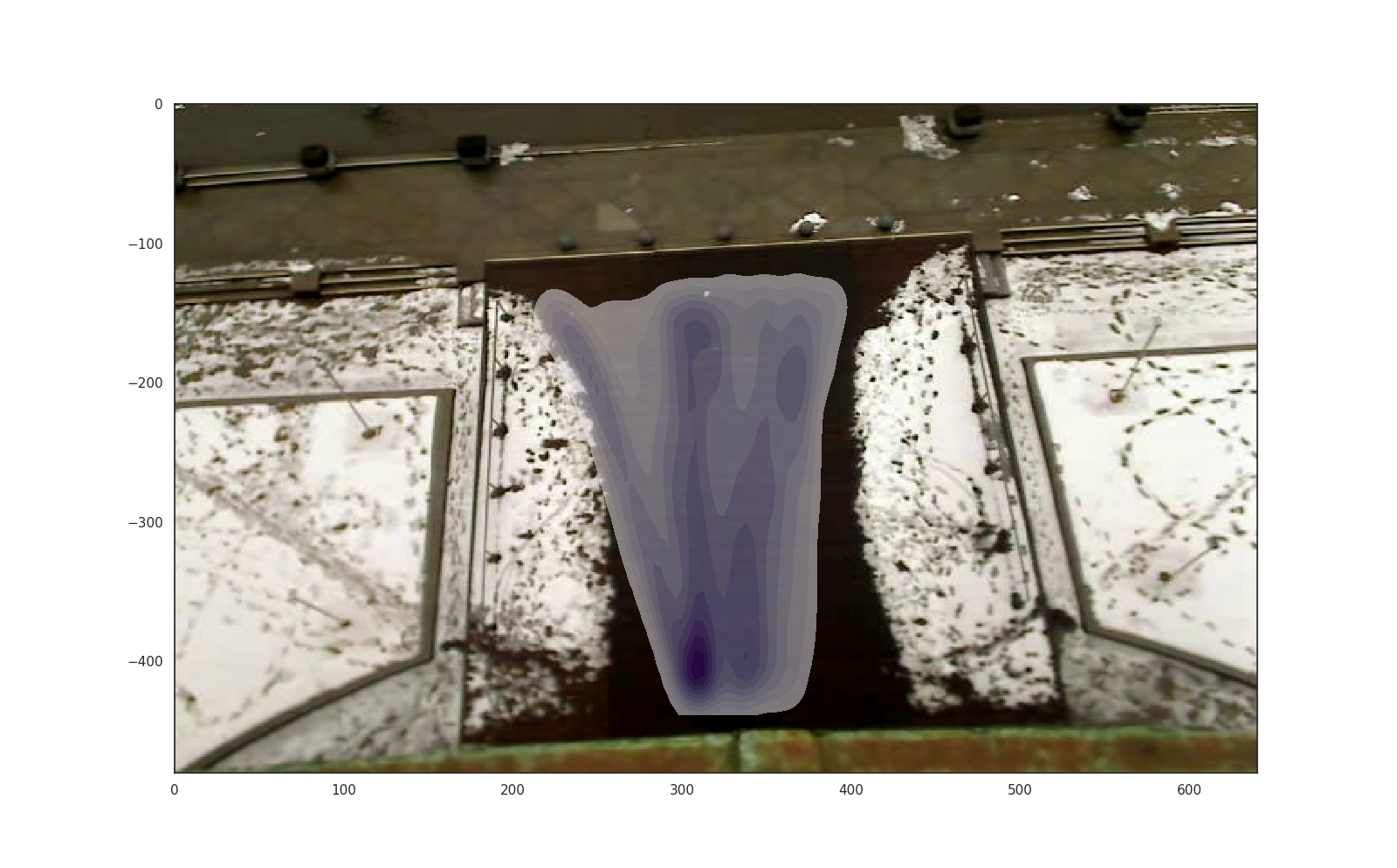}
	\end{subfigure}
	\captionsetup{format=plain, font=footnotesize, labelfont=bf}
	\caption{Trajectory heatmaps: the input data, the generated trajectories, and the final simulated agent motion. }
	\label{fig:figHeatmap}
\end{figure}


\SubsectionShort{Result 2: Trajectories} 
Next, we used our system to generate 352 new trajectories, and we used them to simulate a crowd.
The first two heatmaps in Fig.~\ref{fig:figHeatmap} show that generated trajectories (middle) 
are similarly distributed over the environment as the real data (left). 

The third heatmap shows the final motion of the simulated agents with route following and collision avoidance. 
In this scenario, agents are well capable of following their given trajectories.


\SubsectionShort{Computation time}
We used CUDA to run our GAN on a NVIDIA Quadro M1200 GPU with 4GB of GDDR5 memory. 
With this set-up, generating a batch of 1024 trajectories (with a maximum length of 40 points) took $152$ms, 
meaning that the average generation time was $0.15$ms per trajectory.
Thus, after training, the system is sufficiently fast for real-time insertions of trajectories into a crowd.

\section{Conclusions \& Future Work}

We have presented a data-driven crowd simulation method that uses GANs 
to learn the properties of input trajectories and then generate new trajectories with similar properties. 
Combined with flexible route following that takes temporal information into account, the trajectories can be used in a real-time crowd simulation. 
Our system can be used, for example, to create variants of a scenario with different densities. 
It can easily be combined with other simulation methods, and it allows interactive applications.

In the future, we will perform a thorough analysis of the trajectories produced by our system, and compare them to other algorithms. 
We will also investigate the exact requirements for reliable training.
Furthermore, our system generates trajectories for individuals, assuming that agents do not influence each other's choices. 
As such, it cannot yet model group behavior, and it performs worse in high-density scenarios where agents cannot act independently.
We would like to handle these limitations in future work. 

\begin{acks}
This project was partly funded by EU project CROWDBOT (H2020-ICT-2017-779942).
\end{acks}

\bibliographystyle{acm}
\bibliography{ms}

\end{document}